\documentclass{osa-article}
\usepackage{multirow}

\journal{osac}

\articletype{Research Article}

\usepackage{lineno}
\usepackage{amsmath}

\usepackage{amssymb}
\usepackage[table]{xcolor}
\usepackage{bm}
\begin{document}
\pagestyle{plain}
\title{HWP-synchronous signal induced by misalignment of the sapphire crystal axis}

\author{Yurika Hoshino\authormark{1},
Ryota Takaku\authormark{2,3},
Tomotake Matsumura\authormark{3,4,5,6},
Yuki Sakurai\authormark{3},
Nobuhiko Katayama\authormark{3},
Makoto Tashiro\authormark{1},
Yukikatsu Terada\authormark{1},
Kosuke Sato\authormark{7},
Satoru Katsuda\authormark{1},
Tommaso Ghigna\authormark{3,8,9},
Kunimoto Komatsu\authormark{2},
Shinya Sugiyama\authormark{1}
}

\address{
\authormark{1}Saitama University, 255 Shimo-Okubo, Sakura-ku, Saitama 338-8570, Japan\\
\authormark{2}Okayama University, 3-1-1 Tsushima-naka, Kita-ku, Okayama 700-8530, Japan\\
\authormark{3}Kavli Institute for the Physics and Mathematics of the Universe (WPI), The University of Tokyo, 5-1-5 Kashiwa-no-Ha, Kashiwa, Chiba 277-8583, Japan\\
\authormark{4}Center for Data Driven Discovery (CD3), Kavli Institute for the Physics and Mathematics of the Universe (IPMU), The University of Tokyo, 5-1-5 Kashiwa-no-Ha, Kashiwa, Chiba 277-8583, Japan\\
\authormark{5}ILANCE, CNRS, University of Tokyo International Research Laboratory, Kashiwa, Chiba 277-8582, Japan\\
\authormark{6} Institute of Space and Astronautical Science (ISAS), Japanese Aerospace Exploration Agency (JAXA), 3-1-1 Yoshinodai, Chuo-ku, Sagamihara, Kanagawa 252-5210, Japan\\
\authormark{7}Department of Astrophysics and Atmospheric Sciences,
Kyoto Sangyo University, Kamigamo-motoyama, Kita-ku, Kyoto 603-8555, Japan\\
\authormark{8}International Center for Quantum-field Measurement Systems for Studies of the Universe and Particles (QUP), High Energy Accelerator Research Organization (KEK), 1-1 Oho, Tsukuba, Ibaraki 305-0801, Japan\\
\authormark{9}{Lawrence Berkeley National Laboratory (LBNL), 1 Cyclotron Road, Berkeley, CA 94720, USA}
}

\email{\authormark{*}ryota.takaku@okayama-u.ac.jp}

\begin{abstract}
We investigate the optical response of a rotating sapphire half-wave plate (HWP) with optic axis misalignment, defined as a deviation of the optic axis from the ideal a-cut orientation.
We extend the transfer-matrix formalism for calculating the modulated transmittance of multi-layer wave plates with optic axis misalignment and oblique incidence.
Using this formalism, we compute the modulated response of a rotating HWP with an optic axis misalignment to linearly polarized incident light.
We experimentally validate the model at 90 GHz using sapphire wave plates with intentionally introduced optic axis misalignments of up to $30^{\circ}$.
Additional even harmonics are observed for oblique incidence and optic axis misalignment, with odd harmonics appearing when both effects are present.
The measured modulation signals are in good agreement with the model predictions.
We also evaluate the potential impact of a illustrative small optic axis misalignment of $3^{\circ}$ on precision polarimetry.
For the configuration considered, the model predicts odd-harmonic amplitudes at the percent level for a single-layer HWP and up to 8\% at individual frequencies for a five-layer achromatic HWP.
\end{abstract}

\section{Introduction}
\label{section1}

Wave plates are optical devices widely used for polarimetry.
In particular, half-wave plates (HWP) 
can be fabricated by cutting a slab of birefringent material with a specific thickness and orientation, such that a $\pi$ retardance is induced between two waves propagating with orthogonal polarization states.
A conventional polarimeter combines an HWP with a polarizer and is widely used for precision polarimetry.
This technique is used in a wide range of applications, including Ly-$\alpha$ polarimetry of the solar chromosphere~\cite{watanabe2011ly} and cosmic microwave background (CMB) polarimetry, biomedical imaging for tissues and materials~\cite{hrebesh2009vivo}, and non-invasive blood glucose monitoring~\cite{roy2020non}.
Among these applications, measurements of CMB polarization aimed at probing cosmic inflation require particularly high precision~\cite{LB_PTEP}.
Such measurements require sensitivity to fractional polarization at the level of $\sim10^{-9}$ and polarization angle accuracy of $\sim1$~arcmin\cite{remazeilles2018exploring}.
In recent years, a number of CMB polarimeters have been designed, characterized, and deployed in millimeter-wave telescopes~\cite{johnson2007maxipol,kusaka2014modulation,bryan1modeling,takakura2017performance,2017A&A...599A..34R,Aboobaker_2018,2018JLTP..193..851H,SimonsObservatory:2018koc}.

Linear polarization modulation can be achieved by rotating the HWP.
An ideal rotating HWP modulates incident linear polarization at four times its rotation frequency.
This is particularly useful to suppress noise because a proper choice of the rotation frequency achieves an
up-conversion that helps mitigate the impact of $1/f$ noise.
The polarization modulation properties of HWPs have been extensively studied theoretically~\cite{Savini:06,Matsumura2009,essinger2013transfer}.
However, in real-world applications, the modulated signal is often contaminated by multiple harmonics.
One well-known effect is due to the surface reflection of the HWP that produces a harmonic at two times the HWP rotational frequency 
due to the HWP’s two distinct refractive indices.
Furthermore, various imperfections originating from the rotation mechanism can be a source of the additional synchronous signals, which lead to systematic effects\cite{takakura2017performance,Ghigna2023,DAlessandro2019}.

For precision polarimetry, it is essential to understand the sources of these synchronous signals.
In this work, we identify the mechanism that generates synchronous signals at various harmonics of the HWP rotational frequency. In particular, we analyze the situation where the optic axis of a birefringent crystal deviates from the ideal orientation, which happens when the crystal’s optical axis is misaligned with its intended direction in the plane of the wave plate.
This effect is particularly relevant when the diameter of the HWP is large.
Modern CMB polarization experiments typically employ sapphire millimeter-wave HWPs with diameters of approximately 500 mm\cite{johnson2007maxipol, SimonsObservatory:2018koc,Kusaka2018ABS}. Such systems often employ multi-layer achromatic HWP (AHWP) to achieve broadband polarization modulation. 
Several examples have been reported in the literature~\cite{Pancharatnam1955,Hanany2005,Pisano2006,Savini:06,Matsumura2009,komatsu2019demonstration,komatsu2021design}.
Due to limitations of the fabrication process, 
perfect alignment between the HWP surface and the crystal’s a-plane cannot be achieved during fabrication.

In this paper, we extend the transfer-matrix formalism of Essinger-Hileman~\cite{essinger2013transfer}
to calculate the HWP-synchronous modulation of rotating multi-layer wave plates with optic axis misalignment under oblique incidence.
We then present transmittance measurements of sapphire wave plates specifically fabricated with a controlled misalignment from the ideal a-cut orientation. These measurements validate the formalism and demonstrate HWP-synchronous signals at multiple harmonics of the rotation frequency.
Finally, we discuss the implications of these additional harmonics for single and multi-layer wave plates used in broadband polarimetry.

The remainder of this paper is organized as follows.
In Sec.~\ref{section2}, we introduce the formalism for modeling multi-layer wave plates with arbitrary optic axis orientations and angles of incidence.
In Sec.~\ref{section3}, we describe the experimental setup and measurements using sapphire wave plates with intentionally fabricated with a deviation from the ideal a-cut orientation.
In Sec.~\ref{section4}, we compare the measured synchronous signals with the model predictions.
In Sec.~\ref{section5}, we discuss the origin and implications of the additional HWP-synchronous harmonics, including their impact on single and multi-layer wave plates used for polarimetry.

\begin{figure}[t]
     \centering
    \includegraphics[width=\linewidth] {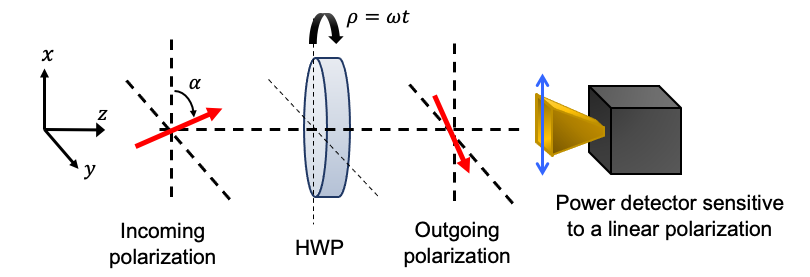}
    \caption{Schematic definition of the optic axis misalignment angle $\psi$. The ideal a-cut HWP has its optic axis in the wave-plate plane ($\psi=0^\circ$), while the c-cut corresponds to $\psi=90^\circ$. The red dotted line indicates the in-plane reference direction for $\psi$.}
    \label{fig:about_modsig}
\end{figure}

\section{HWP polarimetry}
\label{section2}

\subsection{Stokes vector and Mueller matrix formalism}
We follow the formalism described in previous studies~\cite{matsumura2009performance,komatsu2019demonstration}.
The input and output Stokes vectors, $S_{\rm in}$ and $S_{\rm out}$, are related through the Mueller matrix of the wave plate, $\Gamma$. The optical system shown in Fig.~\ref{fig:about_modsig} can be expressed as
\begin{align}
\label{eq:Sout}
S_{\rm out}
=
\Gamma (\rho)
S_{\rm in},
\end{align}
where 
$\rho$ is the HWP rotation angle.
As shown in Fig.~\ref{fig:about_modsig}, for a polarimeter using a detector sensitive to a single linear polarization, the detected intensity can be written as

\begin{align}
\label{eq:IoutwG}
        I_{\rm{out}}&=\frac{1}{2}
    \begin{pmatrix}
        1 & 1 & 0 & 0\\
    \end{pmatrix}S_{\rm{out}}.
\end{align}
To calculate the transmitted intensity, we determine the Mueller matrix $\Gamma$ from the transmission coefficients $t_{pq}$, which is described in Sec.~\ref{sec.transcoeff}.

\subsection{Propagation of EM waves through a HWP for an arbitrary optic axis orientation}

\begin{figure}[tb]
    \centering
        \includegraphics[width=\linewidth] {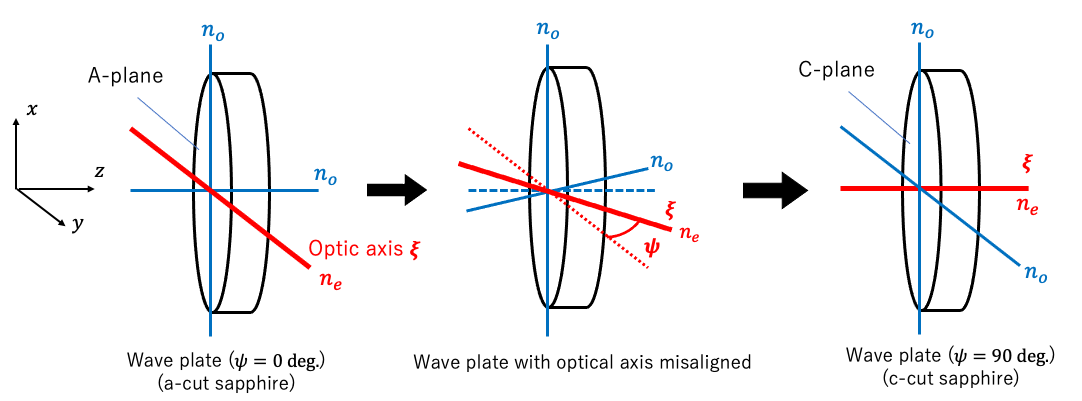}
    \caption{Schematic image of the optic axis misalignment angle $\psi$. }
    \label{fig:about_offaxis}
\end{figure}

For sapphire, the optic axis is the direction associated with the extraordinary refractive index $n_e$.
For an ideal a-cut HWP, the optic axis lies in the $xy$-plane, while for a c-cut plate, it is normal to the $xy$-plane.
We define the misalignment angle $\psi$ such that $\psi = 0^{\circ}$ corresponds to the ideal a-cut and $\psi = 90^{\circ}$ to the c-cut, as illustrated in Fig.~\ref{fig:about_offaxis}.
The HWP thickness $d$ is chosen to provide a retardance of $\Delta\delta = \pi$ at the design frequency, where
$\Delta\delta = 2\pi d |n_e-n_o|\nu/c$,
and $n_o$ and $n_e$ are the refractive indices for the ordinary and extraordinary waves, respectively.
Fabrication limitations can cause the crystal to be cut with the optic axis tilted out of the $xy$-plane.

\begin{figure}[tb]
     \centering
    \includegraphics[width=\textwidth] {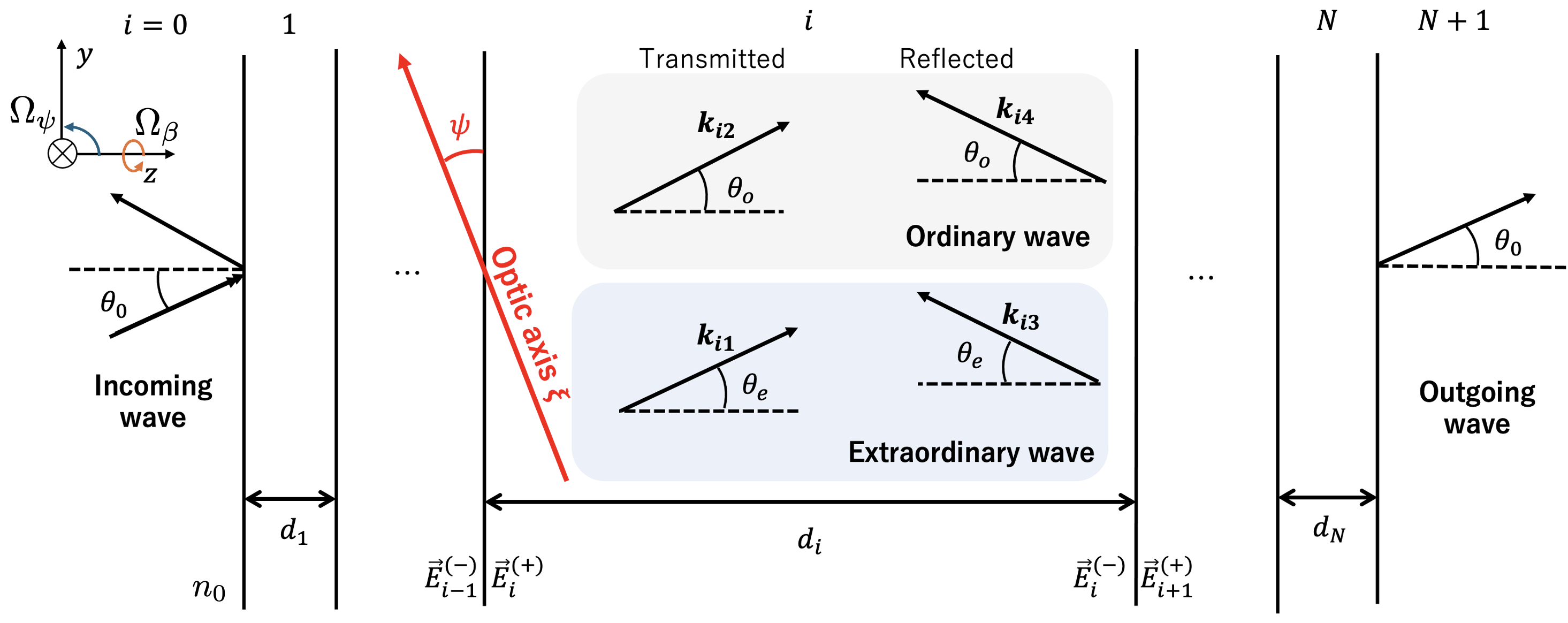}
    \caption{Wave propagation through a birefringent multi-layer medium with an arbitrarily oriented optic axis. The $i=0$ and $i=N+1$ layer are vacuum, while the intermediate layers are birefringent materials.
}
    \label{fig3:offaxis_coordinate}
\end{figure}

The treatment of the multi-layer stack of birefringent materials with arbitrary optic-axis orientations is described in previous studies~\cite{yeh1979electromagnetic, xu2000optical, passler2017generalized}.
For the $i$-th wave plate, we define
\begin{align}
    \beta_i \equiv \rho + \chi_i,
\end{align}
where $\chi_i$ is the relative optic-axis orientation of the $i$-th wave plate.
The unit vector along the optic axis is then written as
\begin{align}
    \hat{\bm{\xi}}_i =
    \begin{pmatrix}
        -\cos{\psi_i}\sin{\beta_i}\\
        \cos{\psi_i}\cos{\beta_i}\\
        \sin{\psi_i}
    \end{pmatrix},
\end{align}
where $\psi_i$ is the optic-axis misalignment angle of the $i$-th wave plate.

The dielectric tensor of the $i$-th wave plate is expressed as
\begin{align}
     \bm{\varepsilon}_i =
     \Omega_{\beta_i}\Omega_{\psi_i}
     \bar{\bm{\varepsilon}}_i
     \Omega_{\psi_i}^{-1}\Omega_{\beta_i}^{-1},
\end{align}
where
\begin{align}
    \Omega_{\beta_i} &=
    \begin{pmatrix}
    \cos{\beta_i} & -\sin{\beta_i} & 0 \\
    \sin{\beta_i} & \cos{\beta_i} & 0 \\
    0 & 0 & 1
    \end{pmatrix},
    \Omega_{\psi_i} =
    \begin{pmatrix}
    1 & 0 & 0 \\
    0 & \cos{\psi_i} & -\sin{\psi_i} \\
    0 & \sin{\psi_i} & \cos{\psi_i}
    \end{pmatrix},
    \bar{\bm{\varepsilon}}_i =
    \begin{pmatrix}
    \epsilon_{o,i} & 0 & 0 \\
    0 & \epsilon_{e,i} & 0 \\
    0 & 0 & \epsilon_{o,i}
    \end{pmatrix}.
\end{align}
The rotation matrix $\Omega_{\beta_i}$ represents the orientation of the $i$-th wave plate about the $z$-axis, including both the HWP rotation $\rho$ and the relative optic-axis orientation $\chi_i$, while $\Omega_{\psi_i}$ represents the optic-axis misalignment about the $x$-axis.
Fig.~\ref{fig3:offaxis_coordinate} shows the cross-sectional view of the HWP in the $yz$-plane.
The incoming wave is incident on the $xy$-plane at an angle $\theta_0$.

At the boundary between the $(i-1)$-th and $i$-th layer, the wave vector $\bm{k}_{ij}$ is given by

\begin{align}
\label{eq:kij}
    \bm{k}_{ij}
    =
    \frac{2\pi\nu}{c}
    \begin{pmatrix}
        0\\
        n_{ij}\sin{\theta_{ij}}\\
        s_j n_{ij}\cos{\theta_{ij}}
    \end{pmatrix},
\end{align}
where $s_j=+1$ for transmitted waves and $s_j=-1$ for reflected waves.
The index $j$ denotes
\begin{align}
\begin{split}
    j=1: \,&\rm{extraordinary\, transmitted \ wave},\\
    j=2: \,&\rm{ordinary\, transmitted \ wave},\\
    j=3: \,&\rm{extraordinary\, reflected \ wave},\\
    j=4: \,&\rm{ordinary\, reflected \ wave}.
\end{split}
\end{align}
The phase matching condition at the boundary gives
\begin{align}
\label{eq:phase}
    n_0\sin{\theta_0}
    =
    n_{1j}\sin{\theta_{1j}}
    =
    \cdots
    =
    n_{ij}\sin{\theta_{ij}},
\end{align}
because the tangential phase must be continuous across the boundary. In Eq.~\ref{eq:phase}, $n_0$ and $\sin{\theta}$ are refractive index and incident angle of the incident layer, respectively.

Here, $\eta$ is the angle between the optic axis $\hat{\bm{\xi}}$ and the propagation direction of the extraordinary wave.
Using the unit direction vector
\begin{align}
\label{eq:khat_ij}
    \hat{\bm{k}}_{ij}
    =
    \begin{pmatrix}
        0\\
        \sin{\theta_{ij}}\\
        s_j\cos{\theta_{ij}}
    \end{pmatrix},
\end{align}
we define
\begin{align}
\label{eq:eta}
    \cos{\eta}
    =
    \hat{\bm{k}}_{ij}\cdot\hat{\bm{\xi}}
    =
    \xi_y\sin{\theta_{ij}}+\xi_z s_j\cos{\theta_{ij}}.
\end{align}

Following \cite{born_wolf_bhatia_clemmow_gabor_stokes_taylor_wayman_wilcock_1999}, the velocity can be written as a function of the refractive index:
\begin{align}
\begin{split}
    v^{\prime} &= \frac{c}{n_o},\\
    v^{\prime\prime}
    &= c\sqrt{
        \frac{\cos^2{\eta}}{n_o^2}
        +
        \frac{\sin^2{\eta}}{n_e^2}
    }.
\end{split}
\end{align}
Thus, the effective refractive index for the four modes is
\begin{align}
\label{eq:n_ij}
\begin{split}
    n_{ij} &= n_o
    \hspace{1cm}
    (j=2,4:\ {\rm ordinary\,wave}),\\
    n_{ij} &=
    \frac{n_o n_e}
    {\sqrt{n_e^2\cos^2{\eta}+n_o^2\sin^2{\eta}}}
    \hspace{1cm}
    (j=1,3:\ {\rm extraordinary\,wave}).
\end{split}
\end{align}
Although in Eq.~\ref{eq:n_ij}, $n_{ij}$ are unknown, Eq.~\ref{eq:phase} allows $\theta_{ij}$, and hence $\eta$, to be expressed as a function of $n_{ij}$.
Substituting this relation into Eq.~\ref{eq:n_ij} yields an equation in which $n_{ij}$ are the only unknown.
We then reduce this equation to a polynomial and solve for $n_{ij}$ for each mode.

\begin{figure}[tb]
     \centering
    \includegraphics[width=\textwidth] {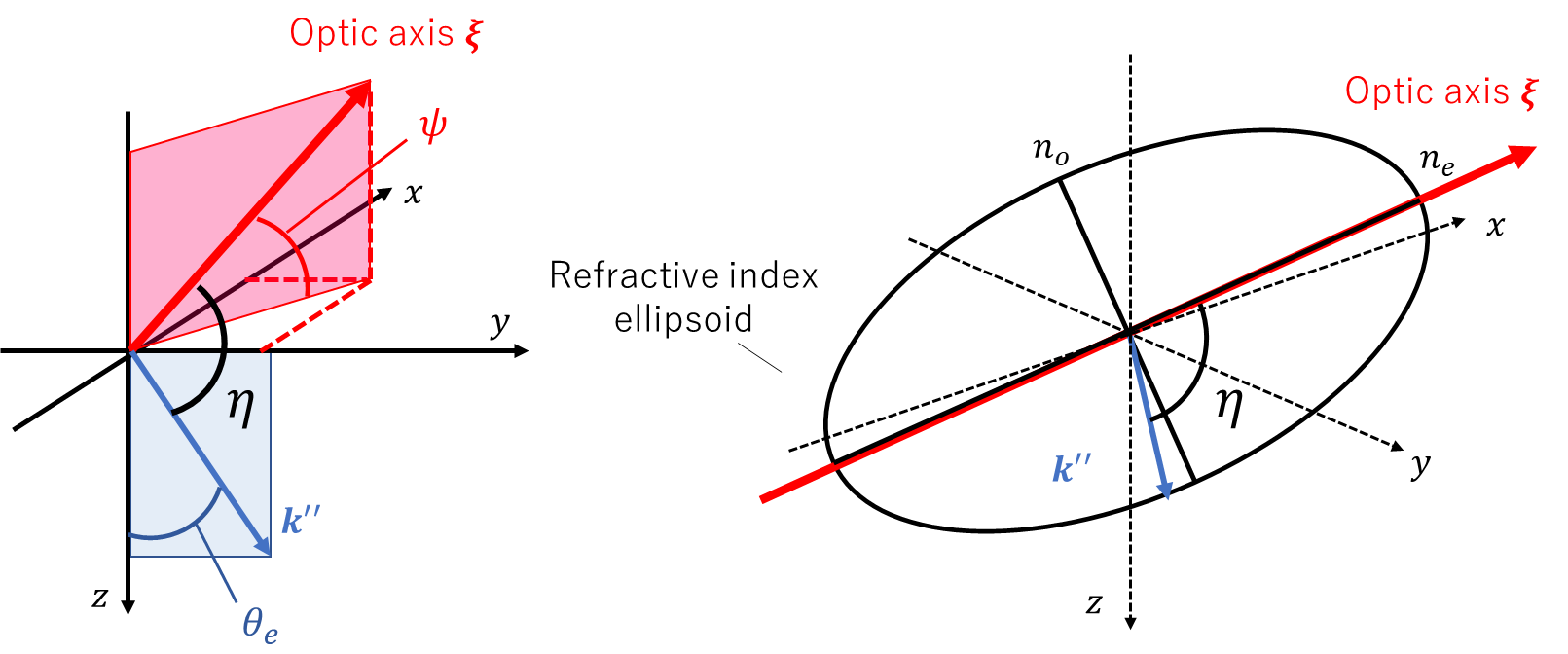}
    \caption{
    Refractive index ellipsoid for the extraordinary wave. The angle $\eta$ is defined between the optic axis $\bm{\xi}$ and the extraordinary-wave vector $\bm{k}_e$.
}
    \label{fig3:about_eta}
\end{figure}

\subsection{Transmission coefficients of a multi-layer wave plate with a misaligned optic axis}
\label{sec.transcoeff}

We model the Mueller matrix of a wave plate with an arbitrarily oriented optic axis following the formalism of Passler et al.~\cite{passler2017generalized}. At the boundary between the $(i-1)$-th and $i$-th layer, the electric field amplitudes are related by the matrices $\bm{A}_{i-1}$ and $\bm{A}_i$ as

\begin{eqnarray}
    \bm{A}_{i-1}\bm{E}_{i-1}^{(-)}
    =
    \bm{A}_{i}\bm{E}_{i}^{(+)} ,
\end{eqnarray}
where
\begin{eqnarray}
\bm{E}_{i}=
\left(
E_{i1},E_{i2},E_{i3},E_{i4}
\right)
\end{eqnarray}
The explicit form of $\bm{A}_i$ is given in Appendix~\ref{appendixA}.
The superscripts $(+)$ and $(-)$ denote the fields at the two boundaries of each layer, as illustrated in Fig.~\ref{fig3:offaxis_coordinate}.
Propagation through the $i$-th birefringent layer is described by the propagation matrix $\bm{P}_i$ as

\begin{align}
    \bm{P}_i
    =
    \begin{pmatrix}
        e^{-i(2\pi\nu/c)q_{i1}d_i} & 0 & 0 & 0 \\
        0 & e^{-i(2\pi\nu/c)q_{i2}d_i} & 0 & 0 \\
        0 & 0 & e^{-i(2\pi\nu/c)q_{i3}d_i} & 0 \\
        0 & 0 & 0 & e^{-i(2\pi\nu/c)q_{i4}d_i}
    \end{pmatrix},
\end{align}
where
\begin{align}
    q_{ij}=s_j n_{ij}\cos\theta_{ij}.
\end{align}
The transfer matrix of the $i$-th layer is then given by
\begin{align}
\label{eq:DefT_sandbox}
    \bm{T}_i
    =
    \bm{A}_i
    \bm{P}_i
    \bm{A}_i^{-1}.
\end{align}
For an $N$-layer system, the total transfer matrix connecting the incident and outgoing layers is
\begin{align}
\label{eq:Mtot_sandbox}
\bm{T}_{\rm tot}
=
\bm{A}_0^{-1}
\bm{T}_1\bm{T}_2\cdots\bm{T}_N
\bm{A}_{N+1}
\end{align}

The field amplitudes on the incident and outgoing sides are related by
$\bm{E}_0 = \bm{T}_{\rm tot}\bm{E}_{N+1}$.
The outgoing field amplitudes can therefore be obtained by inverting $\bm{T}_{\rm tot}$.
After transforming the four-mode amplitudes into the incident and outgoing $p/s$ polarization bases, we obtain the coefficients of Jones matrix as

\begin{align}
\label{eq:four_t_sandbox}
\begin{split}
    \Delta &= T_{22}T_{44}-T_{24}T_{42},\\
    t_{pp} &= \frac{T_{44}}{\Delta},\qquad
    t_{ss} = \frac{T_{22}}{\Delta},\\
    t_{ps} &= \frac{T_{42}}{\Delta},\qquad
    t_{sp} = \frac{T_{24}}{\Delta}.
\end{split}
\end{align}

The explicit relation between the elements of the total transfer matrix and the Jones matrix coefficients is given in Appendix~\ref{appendixA}.
The Jones matrix is then converted to the Mueller matrix by, 

\begin{align}
    \Gamma_{\alpha\beta}
    =
    \frac{1}{2}
    {\rm Tr}
    \left(
        \sigma_{\alpha}
        \bm{J}
        \sigma_{\beta}
        \bm{J}^{\dagger}
    \right),
\end{align}

where $\sigma_{\alpha}$ are the Pauli-like matrices corresponding to the Stokes convention.

\subsection{Signal modulation by a single HWP}
\begin{figure}[tbh]
     \centering
    \includegraphics[width=0.7\textwidth] {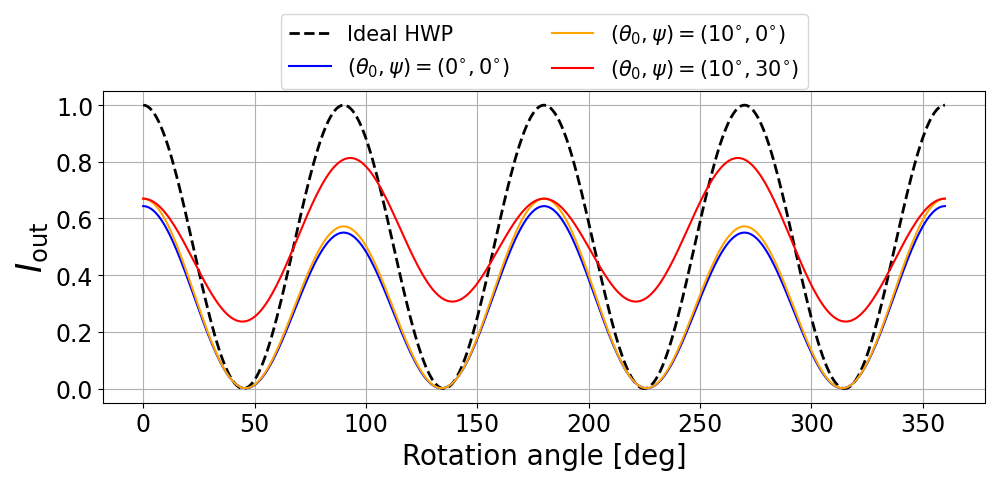}
    \caption{
    Simulation of the modulation signals for different configurations of $(\theta_0,\psi)$ at 90~GHz. The calculation assumes TM-polarized incidence, $(n_o,n_e)=(3.059,3.397)$~\cite{komatsu2019demonstration}, and a wave plate thickness of 5~mm.
    }
    \label{fig3:single_HWP}
\end{figure}

\begin{figure}[tbhp]
    \centering
    \includegraphics[width=\linewidth]{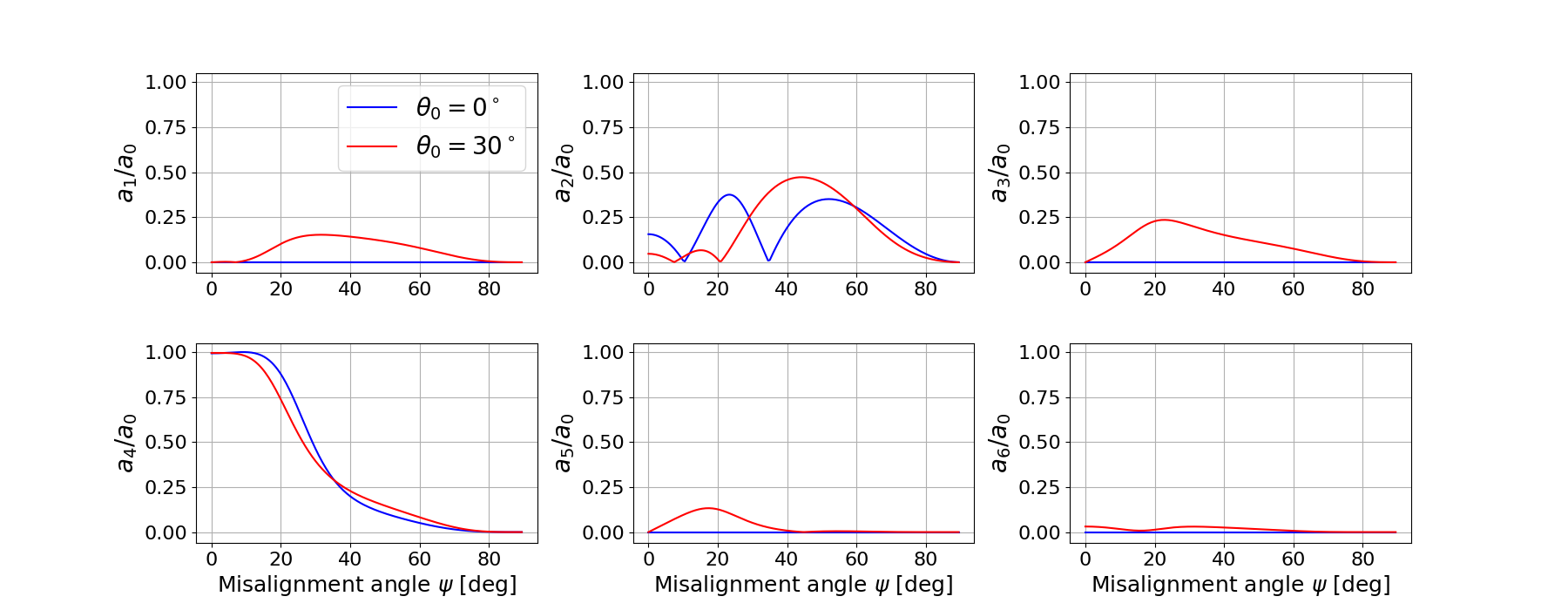}
    \caption{Amplitudes of the first six harmonic terms as a function of the optic axis misalignment angle $\psi$. The blue and red curves correspond to TM-polarized incidence at $0^\circ$ and $30^\circ$, respectively.
    The material properties and frequency are the same as in Fig.~\ref{fig3:single_HWP}.}

    \label{fig3:com_theta0_0degs_30degs}
\end{figure}

The signal modulated by a rotating HWP and detected by a detector sensitive to linearly polarized power can be expressed in a general form as

\begin{align}
  I_{\rm{out}} &= \sum_{m=0}^{l}  a_m \cos{(m\rho + m\phi_{m})} .
 \label{eq:iva_model}
\end{align}
where $a_m$ is the amplitude of each harmonics $m$, $\phi_m$ is the phase of the modulated signal. Examples of simulated signals for a few cases where we vary the incident angle and the misalignment angle are shown in Fig.~\ref{fig3:single_HWP}. In the following sections, to quantify the harmonic components, we fit  the modulated signals with Eq.~\ref{eq:iva_model} up to $l=20$.
We characterize the $m$-th component using the normalized amplitude $a_m/a_0$.
In an ideal case of a rotating HWP, with no reflections at the surface of the plate, we expect only the terms $m=0$ and $m=4$ to be present.
When reflections at the HWP surfaces are not negligible, which is often the case for real-world applications, an additional $m=2$ component appears.
In Fig.~\ref{fig3:com_theta0_0degs_30degs}, we show the relative amplitude of the first six harmonic terms as a function of the misalignement angle for two incident angle values, 0 and 30 degrees.

We find a non-zero $m=6$ component under oblique incidence ($\theta_0\neq 0$) even for $\psi=0^\circ$.
This arises because the effective retardance depends on the rotation angle at oblique incidence.
When $\theta_0 \neq 0^\circ$ and $\psi \neq 0^\circ$, additional odd harmonics such as $m=1$, 3, and 5 also appear.
 \section{Experimental Validation}
\label{section3}
In order to test and validate our framework we prepared a few samples fabricated with the optic axis purposefully misaligned of a known angle between 0 and 30 degrees.

\subsection{Samples}
We prepared four 100 mm diameter flat samples (from Guizhou Haotian Optoelectronics Technology Co., Ltd.), the properties of which are summarized in Table~\ref{tab:Sample_list}. All samples have a design thickness of 5 mm, which provides half-wave retardance at 90~GHz, assuming $n_o = 3.059$ and $n_e = 3.397$~\cite{komatsu2019demonstration}. Sample~1 is designed as an ideal a-cut HWP with no intentional optic axis misalignment. Samples~2--4 are intentionally fabricated with misalignment angles of $\psi = 5^\circ$, $15^\circ$, and $30^\circ$, respectively. The refractive indices listed in the table are obtained from our transmittance measurements following the procedure described in Ref.~\cite{Takaku:21}.
The extraordinary refractive index $n_e$ could not be independently determined for Samples~2--4 with the present measurement method because it relies on the optic axis being properly aligned.
We therefore use the value measured for Sample~1 for Samples~2--4.
This assumption is supported by the measured ordinary refractive indices, which are mutually consistent within their uncertainties for all four samples.

\begin{table}[tb]
\centering
    \caption{Properties of the samples used in the measurements.}
    \begin{tabular}{c|cccc}
    & Sample 1 & Sample 2 & Sample 3 & Sample 4 \\
    \hline
    Misalignment & \multirow{2}{*}{$0\pm0.2$} & \multirow{2}{*}{$5\pm0.5$} & \multirow{2}{*}{$15\pm0.5$} & \multirow{2}{*}{$30\pm0.2$}\\
    angle $\psi$ [deg.] & & & &\\
    \hline
    Thickness $d$ [mm] & $5.002\pm0.004$ & $5.037\pm0.012$ & $5.039\pm0.003$ & $5.019\pm0.002$\\
     \hline
    Refractive index $n_o$ & $3.064\pm0.002$ & $3.062\pm0.007$ & $3.064\pm0.002$ & $3.063\pm0.001$ \\
    Refractive index $n_e$& $3.402\pm0.003$ & - & - & -
    \label{tab:Sample_list}
    \end{tabular}
\end{table}

\subsection{Experimental setup and analysis}
\label{experimental_setup}

\begin{figure}[tb]
    \centering
    \includegraphics[width=\linewidth] {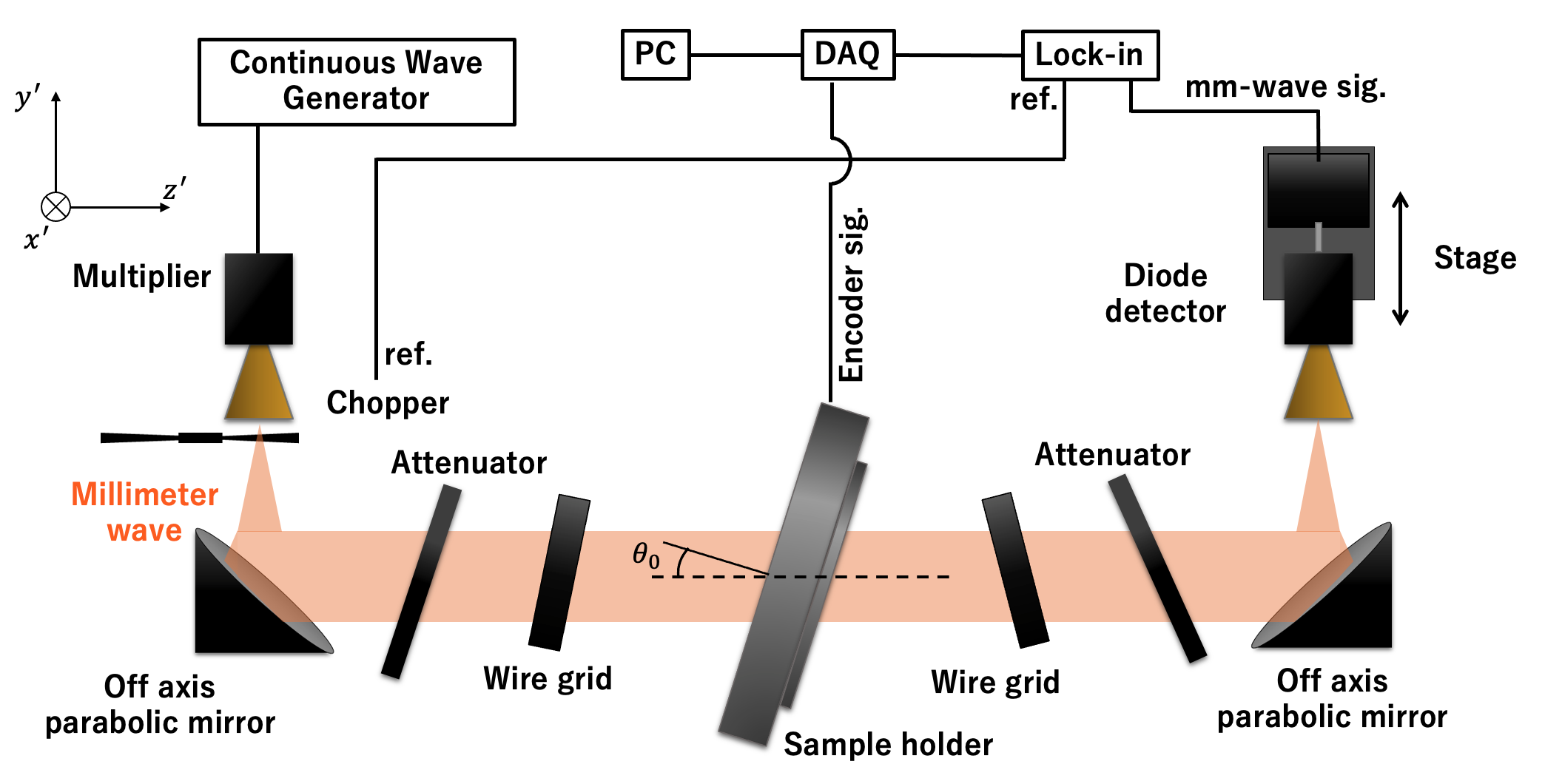}
    \caption{Schematic diagram of the experimental setup.}
    \label{fig4:cw_setup}
\end{figure}
Fig.~\ref{fig4:cw_setup} shows the measurement setup. 
A continuous-wave (CW) source followed by an active $\times 6$ frequency multiplier generates a monochromatic millimeter-wave signal.
The signal is modulated at 200~Hz by an optical chopper.
The signal is then detected by a diode detector, and demodulated by a lock-in amplifier.
The sample is placed between two parabolic mirrors that collimate the transmitted beams. The sample holder allows rotation about the $z'$-axis and tilting about the $x'$-axis by $\theta_0$, thus controlling the angle of incidence. Two wire grids and two attenuators are placed along the optical path to suppress standing waves. Tilting the sample about the $x'$-axis results in TM-polarized incidence. To further reduce standing-wave effects, the detected power is measured at two detector positions separated by a quarter wavelength along the $z$-axis, and the two measurements are averaged. The same procedure is performed with and without the sample. The transmittance is then obtained by taking the ratio of the measured signals with and without the sample at each frequency. This procedure is described in detail in Takaku et al.~\cite{takaku2020jap}.

\subsection{Modulated time-ordered data}
The sample is mounted on a rotary stage driven by a geared DC motor at a rotation frequency of approximately 0.35~Hz. The rotation angle is monitored by an optical encoder mounted on the rotary stage. The encoder consists of 120 arc-shaped slots, each spanning $3^\circ$, together with an LED and photodiode mounted on the stator. The rotary stage is mounted on a rotation stage to adjust the incident angle.

During each measurement, the source frequency is fixed at 90~GHz. The encoder signal and the demodulated optical signal from the lock-in amplifier are recorded synchronously for approximately 20~s at a sampling rate of 1~kHz. The encoder signal is then converted to time-ordered HWP angle data. For each data set, we first identify the maximum of the optical signal near rotation angle $\rho = 0^\circ$. Because the optical encoder provides only the relative rotation angle, and the absolute angle limited by the slot width, we correct for a global angular offset such that the time-ordered data begin at the signal maximum. The optical signal is then binned in $1^\circ$ intervals within 0 to 360 degrees using the encoder-measured HWP angle. Each bin contains approximately 50 samples, and the uncertainty in each bin is estimated from the root-mean-square variation of the samples. The binned data are fitted with Eq.~\ref{eq:iva_model}, including harmonics up to $l=20$.
\section{Results}
\label{section4}
\begin{figure}[t]
    \centering
    \includegraphics[width=\linewidth]{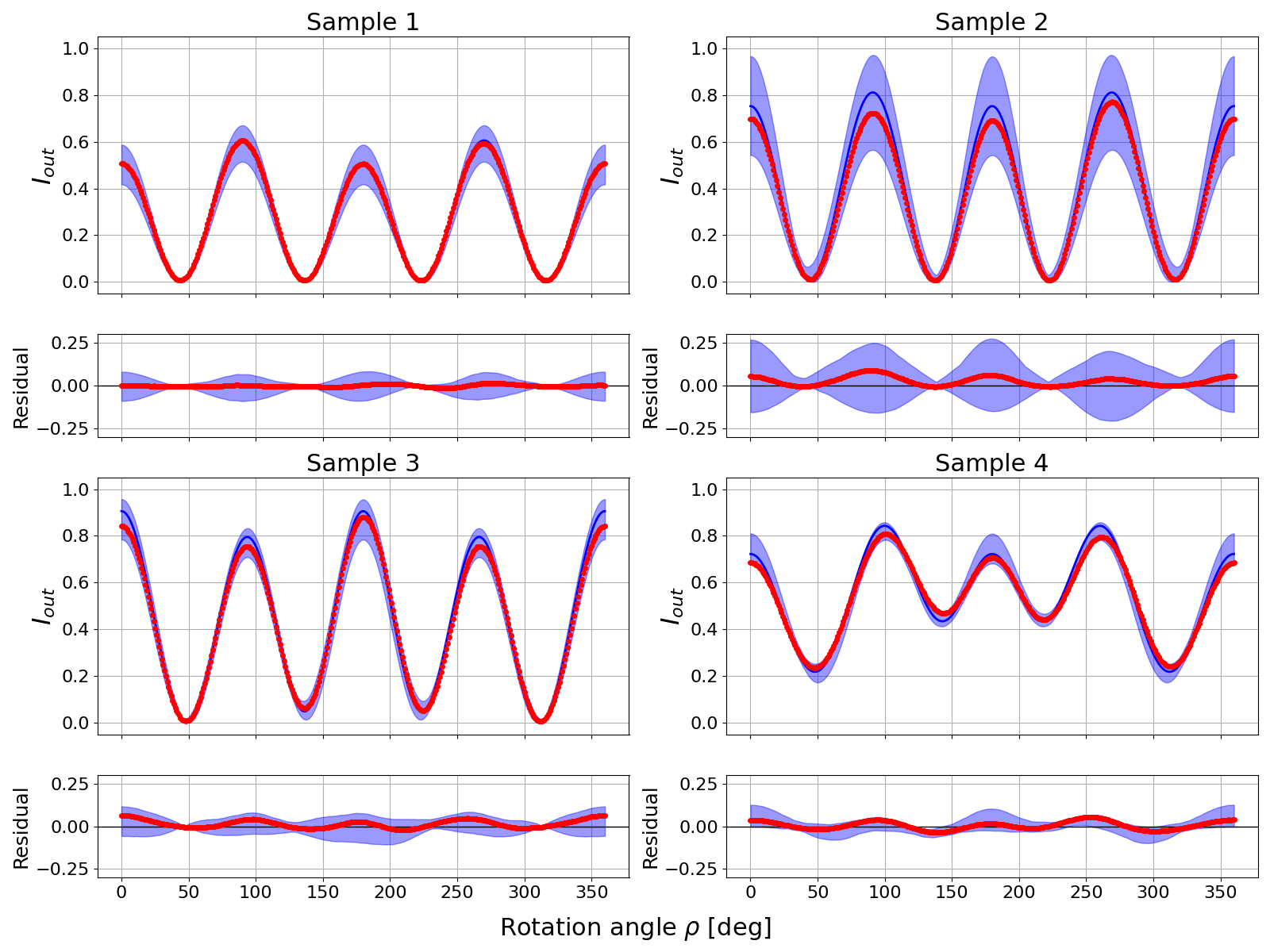}
    \caption{The measured modulation signals with the model predictions for Samples~1--4. The red points show the measurements, the blue solid lines show the predictions, and the shaded regions indicate the $1\sigma$ prediction uncertainties from the sample thickness, refractive indices and optic axis orientation listed in Tab.~\ref{tab:Sample_list}. The lower panels show the absolute values of residuals between the measurements and predictions.
}
    \label{fig4:offaxis_results}
\end{figure}

Fig.~\ref{fig4:offaxis_results} shows the modulated signals measured at 90~GHz for the four samples for a fixed incident angle. For sample 1 we show the data for 0 degree incident angle, while for Samples~2--4 we show the 30 degree case. The predicted curves are calculated using only the sample parameters listed in Tab.~\ref{tab:Sample_list} and the formalism described in Sec.~\ref{section2}, with no free fitting parameters. The measurements are consistent with the predictions within the estimated uncertainties. The RMS of the residuals between the measurements and predictions are 0.006, 0.037, 0.027, and 0.024 for the Samples~1--4, respectively.

\begin{figure}[t]
    \centering
    \includegraphics[width=\linewidth]{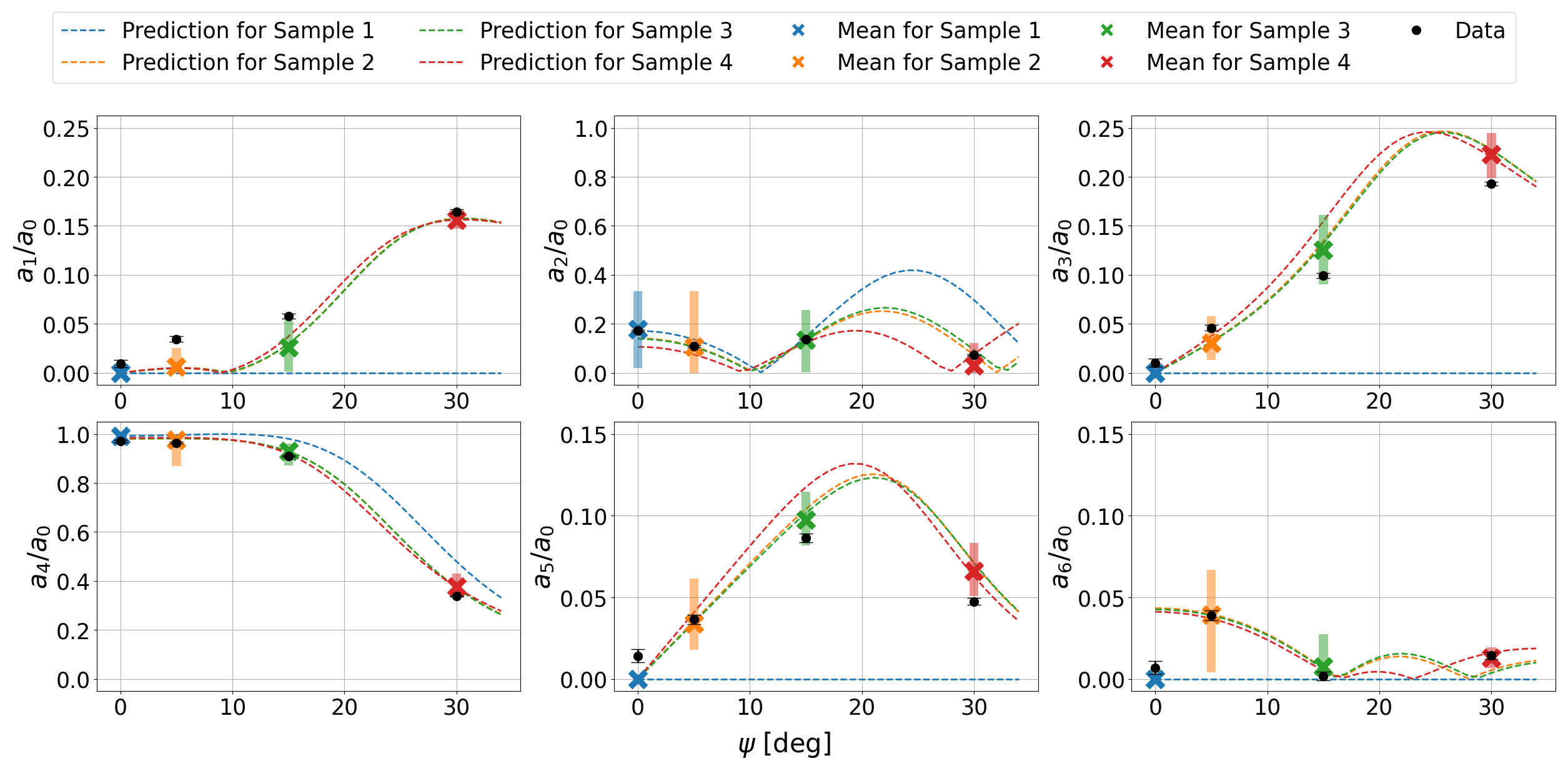}
    \caption{
    The amplitudes of harmonics $a_m/a_0$ obtained from the measurements and model predictions as a function of the optic axis misalignment angle $\psi$. The curves for sample 1 and for sample 2--4 are conducted at 0 degrees and 30 degrees, respectively. The shaded regions indicate the $1\sigma$ prediction uncertainties. Each color corresponds to a different sample, accounting for the differences in sample thickness and refractive indices.}
    \label{fig4:cw_results_parameter}
\end{figure}

\begin{figure}[t]
    \centering
    \includegraphics[width=\linewidth]{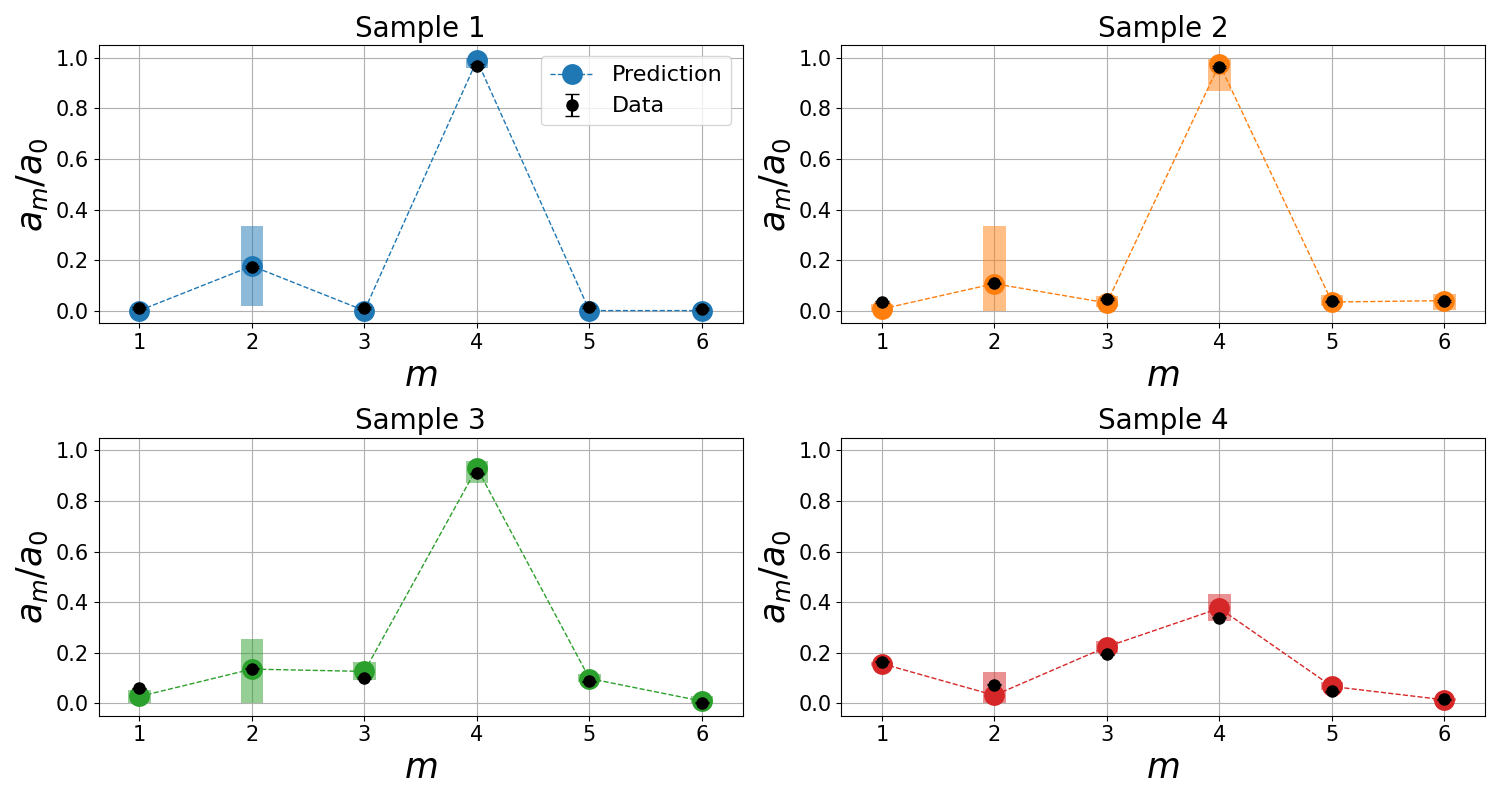}
    \caption{The amplitudes of harmonics $a_m/a_0$ obtained from the measurements and model predictions as a function of harmonics number $m$.
}
    \label{fig4:cw_results_parameter2}
\end{figure}

Fig.~\ref{fig4:cw_results_parameter} shows the normalized harmonic amplitudes $a_m/a_0$ as a function of the optic axis misalignment angle $\psi$. The measured values are compared with the corresponding model predictions for each sample, accounting for the small differences in thickness and refractive indices. Overall, the measurements agree well with the predictions, with most data points lying within the $1\sigma$ uncertainty bands derived from the input parameters. Fig.~\ref{fig4:cw_results_parameter2} shows the same normalized amplitudes as a function of harmonic number $m$. The $m=2$ and $m=4$ components are dominant, while additional non-zero harmonics are observed for samples with $\psi \neq 0^\circ$.

\section{Discussion}
\label{section5}

\subsection{Origin of the $m\ne2, 4$ harmonics}
\label{section5-1}

\begin{figure}
    \centering
    \includegraphics[width=0.45\linewidth]{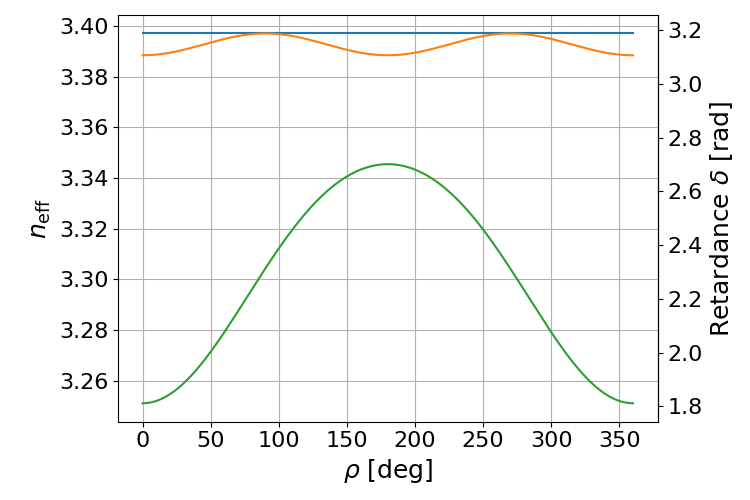}
    \includegraphics[width=0.45\linewidth]{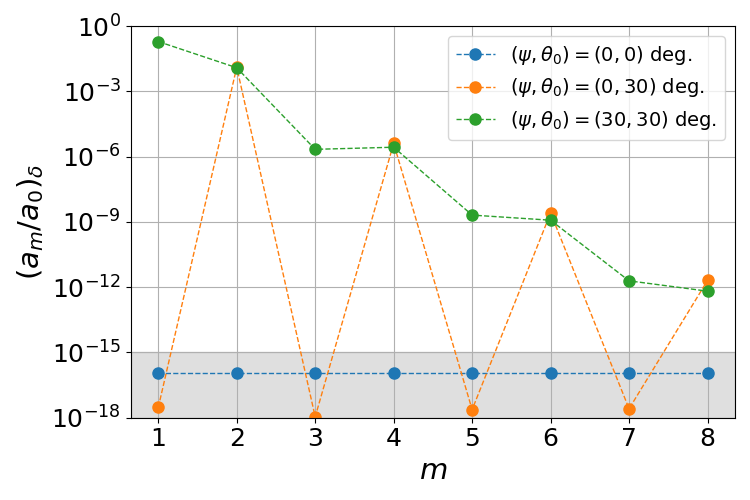}
    \caption{
    The left panel shows the effective refractive index $n_{\rm eff}$ and the corresponding retardance $\delta$ as functions of the HWP rotation angle. The right panel shows the harmonic coefficients obtained by fitting $\delta$ with Eq.~\ref{eq:iva_model}.
    }
    \label{fig:cosdel}
\end{figure}

For an ideal HWP, the modulation is dominated by the $m=2$ and $m=4$ harmonics when reflection is included, whereas we observe non-zero components with $m \ne 2,4$ in Eq.~\ref{eq:iva_model} for $\theta_0 \ne 0$ and/or $\psi \ne 0$.
Fig.~\ref{fig:cosdel} shows the effective refractive index of the extraordinary wave, $n_{\mathrm{eff}} = n_{(i=1,j=1)}$ as defined in Eq.~\ref{eq:n_ij}, for a single wave plate under three representative configurations.
Here, $i=1$ denotes the wave-plate layer and $j=1$ the extraordinary transmitted mode, and $\eta$ in Eq.~\ref{eq:n_ij} is the angle between the optic axis and the propagation direction of the extraordinary wave.
As the HWP rotates, the orientation of the optic axis relative to the propagation direction changes. Under oblique incidence, this makes $\eta$, and hence $n_{\mathrm{eff}}$, dependent on the HWP rotation angle $\rho$.
We define the corresponding effective retardance as
\begin{align}
\delta = 2\pi\frac{\nu}{c}(n_{\mathrm{eff}}-n_o)d.
\end{align}
For normal incidence without misalignment, $n_{\mathrm{eff}}$ is independent of the HWP rotation angle, and the retardance is also constant.
In contrast, for $(\theta_0,\psi)=(30^\circ,0^\circ)$ and $(30^\circ,30^\circ)$, $n_{\mathrm{eff}}$ varies with the HWP rotation angle.
This rotation-angle-dependent retardance generates additional harmonics in the modulated signal, leading to non-zero amplitudes at $m\ne2,4$ in Eq.~\ref{eq:iva_model}.

\subsection{Model and Uncertainty}
\label{section5-2}
We evaluate how the HWP parameter uncertainties propagate to the model predictions and how these compare to the measurements. The \(1\sigma\) uncertainties of the sample parameters summarized in Table~\ref{tab:Sample_list} are propagated to the predicted modulation signals. The resulting uncertainty is shown in Fig.~\ref{fig4:offaxis_results}. The measurement uncertainties are shown as error bars on the data points.
The prediction uncertainties include contributions from the sample thickness, refractive indices, and optic axis orientation.
The uncertainty for Sample~2 is larger than those for the other samples, primarily because of its larger thickness uncertainty.
Within these estimated uncertainties, the measurements are consistent with the model predictions.

\subsection{Implications of optic-axis misalignment for HWP polarimetry}
\label{section5-3}
\begin{figure}[t]
    \centering
    \includegraphics[width=\linewidth]{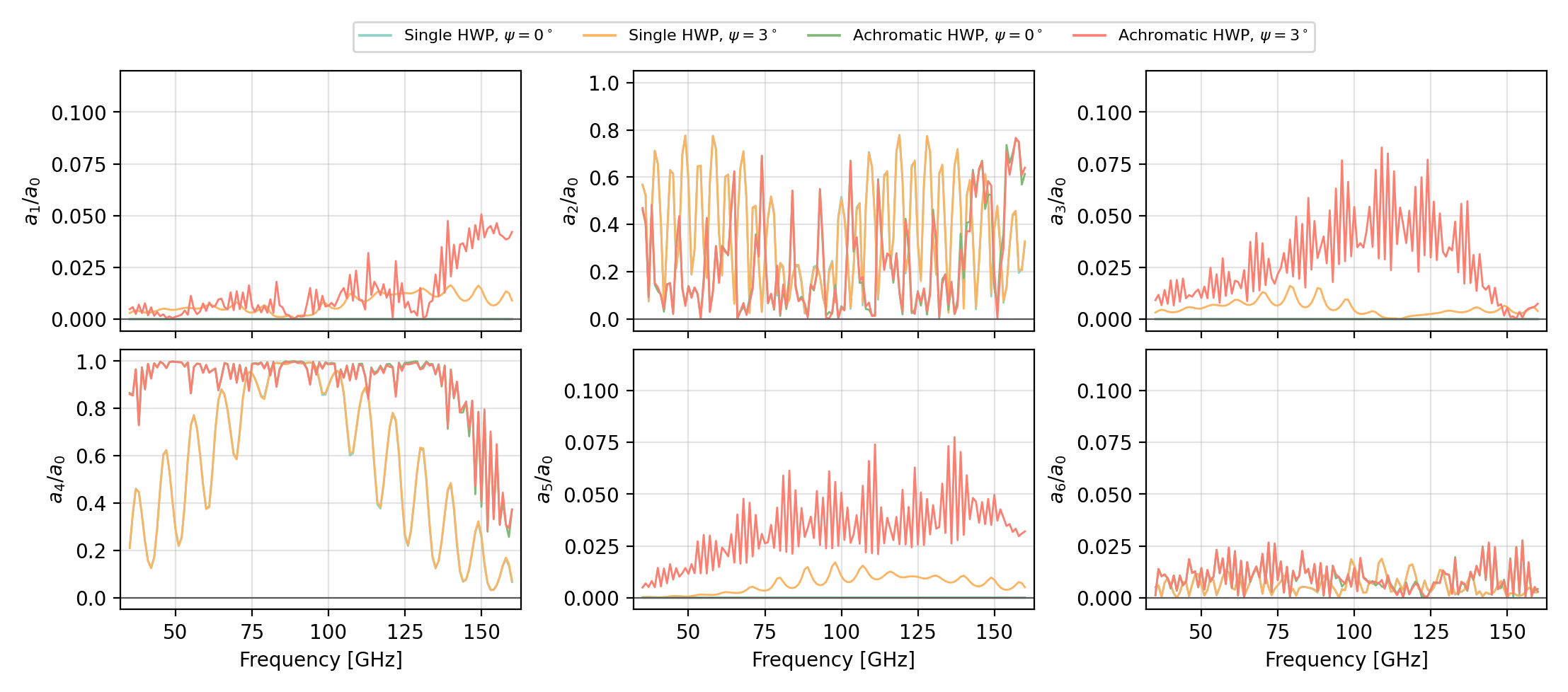}
    \caption{
    Frequency dependence of the amplitudes of harmonics for a single-layer HWP and a five-layer AHWP with optic axis misalignment. The calculations assume a plate thickness of 5~mm, an incident angle of $15^\circ$, and $(n_o,n_e)=(3.059,3.397)$.
    }
    \label{fig11:AHWPpred}
\end{figure}

As an illustrative case, we evaluate the impact of an optic-axis misalignment of $\psi = 3^\circ$ on polarimetric measurements. Fig.~\ref{fig11:AHWPpred} shows the fractional amplitudes of the harmonic components as a function of frequency for a single-layer HWP and a five-layer AHWP, for both $\psi = 0^\circ$ and $3^\circ$. As a representative oblique-incidence case for a wide-field CMB polarimeter, we adopt $\theta_0=15^\circ$, which is comparable to incidence angles considered for wide-field CMB telescopes~\cite{10.1093/mnras/stab317}.
For the five-layer AHWP, we adopt the anti-symmetric design centered at 90 GHz described in Komatsu et al.~\cite{komatsu2021design}.
For $\psi = 3^\circ$, odd-harmonic components at the percent level appear in the single-layer HWP and can reach several percent in the five-layer AHWP.
In the configuration considered here, the odd-harmonic amplitudes reach up to approximately 8\% for the five-layer AHWP.
When \(a_m/a_0\) is uniformly averaged over 34--161 GHz, the odd-harmonic amplitudes remain below 1\% for the single-layer HWP and reach up to approximately 3.2\% for the five-layer AHWP.
The relevance of these additional harmonics depends on the requirements of the specific polarimetric application, but they should be considered when designing broadband HWP polarimeters.

\section{Conclusions}
\label{sec:conslusions}
We investigated the effect of optic axis misalignment on the synchronous signal of a rotating sapphire HWP.
We applied a transfer-matrix formalism to calculate the transmission and modulation properties of multi-layer wave plates with arbitrary optic axis and incident angle orientations. The model was validated experimentally at 90~GHz using sapphire wave plates with intentional optic axis misalignments.
The measured modulation signals are found to be consistent with the model predictions within the experimental uncertainties.
We find that oblique incidence and optic axis misalignment generate additional HWP synchronous harmonic components beyond the commonly expected $m=2$ and $m=4$ terms.
In particular, odd harmonics appear when optic axis misalignment is combined with oblique incidence.
For the illustrative case of \(\psi=3^\circ\), the odd-harmonic amplitudes reach several percent at individual frequencies for a five-layer AHWP and remain at the few-percent level when \(a_m/a_0\) is averaged over 34--161 GHz. 
These effects should therefore be taken into account when designing HWP-based precision polarimeters.

\section*{Funding}
This work was supported by JSPS KAKENHI Grant Number JP24K17064, 23H00107.
This work is performed in part at Center for Data-Driven Discovery, Kavli IPMU (WPI). The Kavli IPMU is supported by World Premier International Research Center Initiative (WPI Initiative), MEXT, Japan.

RT acknowledges the RECTOR program in Okayama University.

\section*{Disclosures}
The authors declare no conflicts of interest.

\section*{Data Availability}
The simulation code used in this work is publicly available at https://github.com/ipmu-lab/offaxsim.
Data underlying the results presented in this paper are available from the corresponding author upon reasonable request.

\appendix
\setcounter{section}{0}
\renewcommand{\thesection}{\Alph{section}}
\setcounter{equation}{0}
\renewcommand{\theequation}{\Alph{section}.\arabic{equation}}
\setcounter{figure}{0}
\renewcommand{\thefigure}{\Alph{section}.\arabic{figure}}
\setcounter{table}{0}
\renewcommand{\thetable}{\Alph{section}.\arabic{table}}

\bibliography{thesis}

@book{born_wolf_bhatia_clemmow_gabor_stokes_taylor_wayman_wilcock_1999, 
place={Cambridge}, 
edition={7}, 
title={Principles of Optics: Electromagnetic Theory of Propagation, Interference and Diffraction of Light}, 
DOI={10.1017/CBO9781139644181}, 
publisher={Cambridge University Press}, 
author={Born, Max and Wolf, Emil and Bhatia, A. B. and Clemmow, P. C. and Gabor, D. and Stokes, A. R. and Taylor, A. M. and Wayman, P. A. and Wilcock, W. L.}, 
year={1999}
}

@article{Savini:06,
author = {Giorgio Savini and Giampaolo Pisano and Peter A. R. Ade},
journal = {Appl. Opt.},
number = {35},
pages = {8907--8915},
publisher = {Optica Publishing Group},
title = {Achromatic half-wave plate for submillimeter instruments in cosmic microwave background astronomy: modeling and simulation},
volume = {45},
month = {Dec},
year = {2006},
url = {https://opg.optica.org/ao/abstract.cfm?URI=ao-45-35-8907},
doi = {10.1364/AO.45.008907},
}

@ARTICLE{2018JLTP..193..851H,
       author = {{Hill}, C.~A. and {Kusaka}, A. and {Barton}, P. and {Bixler}, B. and {Droster}, A.~G. and {Flament}, M. and {Ganjam}, S. and {Jadbabaie}, A. and {Jeong}, O. and {Lee}, A.~T. and {Madurowicz}, A. and {Matsuda}, F.~T. and {Matsumura}, T. and {Rutkowski}, A. and {Sakurai}, Y. and {Sponseller}, D.~R. and {Suzuki}, A. and {Tat}, R.},
        title = "{A Large-Diameter Cryogenic Rotation Stage for Half-Wave Plate Polarization Modulation on the POLARBEAR-2 Experiment}",
      journal = {Journal of Low Temperature Physics},
         year = 2018,
        month = dec,
       volume = {193},
       number = {5-6},
        pages = {851-859},
          doi = {10.1007/s10909-018-1980-6},
archivePrefix = {arXiv},
       eprint = {1805.10403},
 primaryClass = {astro-ph.IM},
       adsurl = {https://ui.adsabs.harvard.edu/abs/2018JLTP..193..851H}
}

@ARTICLE{2017A&A...599A..34R,
       author = {{Ritacco}, A. and {Ponthieu}, N. and {Catalano}, A. and {Adam}, R. and {Ade}, P. and {Andr{\'e}}, P. and {Beelen}, A. and {Beno{\^\i}t}, A. and {Bideaud}, A. and {Billot}, N. and {Bourrion}, O. and {Calvo}, M. and {Coiffard}, G. and {Comis}, B. and {D{\'e}sert}, F. -X. and {Doyle}, S. and {Goupy}, J. and {Kramer}, C. and {Leclercq}, S. and {Mac{\'\i}as-P{\'e}rez}, J.~F. and {Mauskopf}, P. and {Maury}, A. and {Mayet}, F. and {Monfardini}, A. and {Pajot}, F. and {Pascale}, E. and {Perotto}, L. and {Pisano}, G. and {Rebolo-Iglesias}, M. and {Rev{\'e}ret}, V. and {Rodriguez}, L. and {Romero}, C. and {Ruppin}, F. and {Savini}, G. and {Schuster}, K. and {Sievers}, A. and {Thum}, C. and {Triqueneaux}, S. and {Tucker}, C. and {Zylka}, R.},
        title = "{Polarimetry at millimeter wavelengths with the NIKA camera: calibration and performance}",
      journal = {\aap},
         year = 2017,
        month = mar,
       volume = {599},
          eid = {A34},
        pages = {A34},
          doi = {10.1051/0004-6361/201629666},
archivePrefix = {arXiv},
       eprint = {1609.02042},
 primaryClass = {astro-ph.IM},
       adsurl = {https://ui.adsabs.harvard.edu/abs/2017A&A...599A..34R}
}

@article{Aboobaker_2018,
doi = {10.3847/1538-4365/aae434},
url = {https://dx.doi.org/10.3847/1538-4365/aae434},
year = {2018},
month = {nov},
publisher = {The American Astronomical Society},
volume = {239},
number = {1},
pages = {7},
author = {The EBEX Collaboration and Asad M. Aboobaker and Peter Ade and Derek Araujo and François Aubin and Carlo Baccigalupi and Chaoyun Bao and Daniel Chapman and Joy Didier and Matt Dobbs and Christopher Geach and Will Grainger and Shaul Hanany and Kyle Helson and Seth Hillbrand and Johannes Hubmayr and Andrew Jaffe and Bradley Johnson and Terry Jones and Jeff Klein and Andrei Korotkov and Adrian Lee and Lorne Levinson and Michele Limon and Kevin MacDermid and Tomotake Matsumura and Amber D. Miller and Michael Milligan and Kate Raach and Britt Reichborn-Kjennerud and Ilan Sagiv and Giorgio Savini and Locke Spencer and Carole Tucker and Gregory S. Tucker and Benjamin Westbrook and Karl Young and Kyle Zilic},
title = {The EBEX Balloon-borne Experiment—Optics, Receiver, and Polarimetry},
journal = {The Astrophysical Journal Supplement Series}
}

@inproceedings{watanabe2011ly,
  title={Ly-alpha polarimeter design for CLASP rocket experiment},
  author={Watanabe, H and Narukage, N and Kubo, M and Ishikawa, R and Bando, T and Kano, R and Tsuneta, S and Kobayashi, K and Ichimoto, K and Trujillo-Bueno, J},
  booktitle={Solar Physics and Space Weather Instrumentation IV},
  volume={8148},
  pages={288--297},
  year={2011},
  organization={SPIE}
}

@article{hrebesh2009vivo,
  title={In vivo imaging of dynamic biological specimen by real-time single-shot full-field optical coherence tomography},
  author={Hrebesh, Molly Subhash and Dabu, Razvan and Sato, Manabu},
  journal={Optics Communications},
  volume={282},
  number={4},
  pages={674--683},
  year={2009},
  publisher={Elsevier}
}

@article{roy2020non,
  title={Non-invasive tracking of polarization rotation from speckle contrast using uncorrelated speckle patterns},
  author={Roy, Abhijit and Singh, Rakesh Kumar and Brundavanam, Maruthi M},
  journal={Journal of Optics},
  volume={22},
  number={12},
  pages={125603},
  year={2020},
  publisher={IOP Publishing}
}

@article{yeh1979electromagnetic,
  title={Electromagnetic propagation in birefringent layered media},
  author={Yeh, Pochi},
  journal={Josa},
  volume={69},
  number={5},
  pages={742--756},
  year={1979},
  publisher={Optical Society of America}
}

@article{passler2017generalized,
  title={Generalized 4$\times$ 4 matrix formalism for light propagation in anisotropic stratified media: study of surface phonon polaritons in polar dielectric heterostructures},
  author={Passler, Nikolai Christian and Paarmann, Alexander},
  journal={JOSA B},
  volume={34},
  number={10},
  pages={2128--2139},
  year={2017},
  publisher={Optical Society of America}
}

@article{essinger2013transfer,
  title={Transfer matrix for treating stratified media including birefringent crystals},
  author={Essinger-Hileman, Thomas},
  journal={Applied Optics},
  volume={52},
  number={2},
  pages={212--218},
  year={2013},
  publisher={Optical Society of America}
}

@article{matsumura2009performance,
  title={Performance of three-and five-stack achromatic half-wave plates at millimeter wavelengths},
  author={Matsumura, Tomotake and Hanany, Shaul and Ade, Peter and Johnson, Bradley R and Jones, Terry J and Jonnalagadda, Prashanth and Savini, Giorgio},
  journal={Applied Optics},
  volume={48},
  number={19},
  pages={3614--3625},
  year={2009},
  publisher={Optical Society of America}
}

@article{komatsu2019demonstration,
  title={Demonstration of the broadband half-wave plate using the nine-layer sapphire for the cosmic microwave background polarization experiment},
  author={Komatsu, Kunimoto and Matsumura, Tomotake and Imada, Hiroaki and Ishino, Hirokazu and Katayama, Nobuhiko and Sakurai, Yuki},
  journal={Journal of Astronomical Telescopes, Instruments, and Systems},
  volume={5},
  number={4},
  pages={044008},
  year={2019},
  publisher={International Society for Optics and Photonics}
}

@article{komatsu2021design,
  title={Design of a frequency-independent optic axis Pancharatnam-based achromatic half-wave plate},
  author={Komatsu, Kunimoto and Ishino, Hirokazu and Katayama, Nobuhiko and Matsumura, Tomotake and Sakurai, Yuki},
  journal={Journal of Astronomical Telescopes, Instruments, and Systems},
  volume={7},
  number={3},
  pages={034005--034005},
  year={2021},
  publisher={Society of Photo-Optical Instrumentation Engineers}
}

@article{xu2000optical,
  title={Optical degeneracies in anisotropic layered media: Treatment of singularities in a 4$\times$ 4 matrix formalism},
  author={Xu, W and Wood, LT and Golding, TD},
  journal={Physical Review B},
  volume={61},
  number={3},
  pages={1740},
  year={2000},
  publisher={APS}
}

@article{bryan1modeling,
author = {Sean A. Bryan and Thomas E. Montroy and John E. Ruhl},
journal = {Appl. Opt.},
number = {32},
pages = {6313--6323},
publisher = {OSA},
title = {Modeling dielectric half-wave plates for cosmic microwave background polarimetry using a Mueller matrix formalism},
volume = {49},
month = {Nov},
year = {2010},
url = {http://opg.optica.org/ao/abstract.cfm?URI=ao-49-32-6313},
doi = {10.1364/AO.49.006313},
}

@article{johnson2007maxipol,
  title={MAXIPOL: cosmic microwave background polarimetry using a rotating half-wave plate},
  author={Johnson, BR and Collins, J and Abroe, ME and Ade, Peter AR and Bock, J and Borrill, J and Boscaleri, A and De Bernardis, P and Hanany, Shaul and Jaffe, AH and others},
  journal={The Astrophysical Journal},
  volume={665},
  number={1},
  pages={42},
  year={2007},
  publisher={IOP Publishing}
}

@article{kusaka2014modulation,
  title={Modulation of cosmic microwave background polarization with a warm rapidly rotating half-wave plate on the Atacama B-Mode Search instrument},
  author={Kusaka, A and Essinger-Hileman, T and Appel, JW and Gallardo, P and Irwin, KD and Jarosik, N and Nolta, MR and Page, LA and Parker, LP and Raghunathan, S and others},
  journal={Review of Scientific Instruments},
  volume={85},
  number={2},
  pages={024501},
  year={2014},
  publisher={American Institute of Physics}
}

@article{takakura2017performance,
  title={Performance of a continuously rotating half-wave plate on the POLARBEAR telescope},
  author={Takakura, Satoru and Aguilar, Mario and Akiba, Yoshiki and Arnold, Kam and Baccigalupi, Carlo and Barron, Darcy and Beckman, Shawn and Boettger, David and Borrill, Julian and Chapman, Scott and others},
  journal={Journal of Cosmology and Astroparticle Physics},
  volume={2017},
  number={05},
  pages={008},
  year={2017},
  publisher={IOP Publishing}
}

@article{remazeilles2018exploring,
  title={Exploring cosmic origins with CORE: B-mode component separation},
  author={Remazeilles, Mathieu and Banday, Anthony J and Baccigalupi, Carlo and Basak, S and Bonaldi, A and De Zotti, G and Delabrouille, J and Dickinson, C and Eriksen, HK and Errard, J and others},
  journal={Journal of Cosmology and Astroparticle Physics},
  volume={2018},
  number={04},
  pages={023},
  year={2018},
  publisher={IOP Publishing}
}

@article{takaku2020jap,
author = {Takaku,R.  and Hanany,S.  and Imada,H.  and Ishino,H.  and Katayama,N.  and Komatsu,K.  and Konishi,K.  and Kuwata-Gonokami,M.  and Matsumura,T.  and Mitsuda,K.  and Sakurai,H.  and Sakurai,Y.  and Wen,Q.  and Yamasaki,N. Y.  and Young,K.  and Yumoto,J. },
title = {Broadband, millimeter-wave anti-reflective structures on sapphire ablated with femto-second laser},
journal = {Journal of Applied Physics},
volume = {128},
number = {22},
pages = {225302},
year = {2020},
doi = {10.1063/5.0022765},
URL = {https://doi.org/10.1063/5.0022765},
eprint = {https://doi.org/10.1063/5.0022765}
}

@article{LB_PTEP,
    author = {LiteBIRD Collaboration and Allys, E and Arnold, K and Aumont, J and Aurlien, R and Azzoni, S and Baccigalupi, C and Banday, A J and Banerji, R and Barreiro, R B and Bartolo, N and Bautista, L and Beck, D and Beckman, S and Bersanelli, M and Boulanger, F and Brilenkov, M and Bucher, M and Calabrese, E and Campeti, P and Carones, A and Casas, F J and Catalano, A and Chan, V and Cheung, K and Chinone, Y and Clark, S E and Columbro, F and D’Alessandro, G and de Bernardis, P and de Haan, T and de la Hoz, E and De Petris, M and Torre, S Della and Diego-Palazuelos, P and Dobbs, M and Dotani, T and Duval, J M and Elleflot, T and Eriksen, H K and Errard, J and Essinger-Hileman, T and Finelli, F and Flauger, R and Franceschet, C and Fuskeland, U and Galloway, M and Ganga, K and Gerbino, M and Gervasi, M and Génova-Santos, R T and Ghigna, T and Giardiello, S and Gjerløw, E and Grain, J and Grupp, F and Gruppuso, A and Gudmundsson, J E and Halverson, N W and Hargrave, P and Hasebe, T and Hasegawa, M and Hazumi, M and Henrot-Versillé, S and Hensley, B and Hergt, L T and Herman, D and Hivon, E and Hlozek, R A and Hornsby, A L and Hoshino, Y and Hubmayr, J and Ichiki, K and Iida, T and Imada, H and Ishino, H and Jaehnig, G and Katayama, N and Kato, A and Keskitalo, R and Kisner, T and Kobayashi, Y and Kogut, A and Kohri, K and Komatsu, E and Komatsu, K and Konishi, K and Krachmalnicoff, N and Kuo, C L and Lamagna, L and Lattanzi, M and Lee, A T and Leloup, C and Levrier, F and Linder, E and Luzzi, G and Macias-Perez, J and Maciaszek, T and Maffei, B and Maino, D and Mandelli, S and Martínez-González, E and Masi, S and Massa, M and Matarrese, S and Matsuda, F T and Matsumura, T and Mele, L and Migliaccio, M and Minami, Y and Moggi, A and Montgomery, J and Montier, L and Morgante, G and Mot, B and Nagano, Y and Nagasaki, T and Nagata, R and Nakano, R and Namikawa, T and Nati, F and Natoli, P and Nerval, S and Noviello, F and Odagiri, K and Oguri, S and Ohsaki, H and Pagano, L and Paiella, A and Paoletti, D and Passerini, A and Patanchon, G and Piacentini, F and Piat, M and Pisano, G and Polenta, G and Poletti, D and Prouvé, T and Puglisi, G and Rambaud, D and Raum, C and Realini, S and Reinecke, M and Remazeilles, M and Ritacco, A and Roudil, G and Rubino-Martin, J A and Russell, M and Sakurai, H and Sakurai, Y and Sasaki, M and Scott, D and Sekimoto, Y and Shinozaki, K and Shiraishi, M and Shirron, P and Signorelli, G and Spinella, F and Stever, S and Stompor, R and Sugiyama, S and Sullivan, R M and Suzuki, A and Svalheim, T L and Switzer, E and Takaku, R and Takakura, H and Takase, Y and Tartari, A and Terao, Y and Thermeau, J and Thommesen, H and Thompson, K L and Tomasi, M and Tominaga, M and Tristram, M and Tsuji, M and Tsujimoto, M and Vacher, L and Vielva, P and Vittorio, N and Wang, W and Watanuki, K and Wehus, I K and Weller, J and Westbrook, B and Wilms, J and Winter, B and Wollack, E J and Yumoto, J and Zannoni, M},
    title = {Probing cosmic inflation with the LiteBIRD cosmic microwave background polarization survey},
    journal = {Progress of Theoretical and Experimental Physics},
    volume = {2023},
    number = {4},
    pages = {042F01},
    year = {2023},
    month = {04},
    issn = {2050-3911},
    doi = {10.1093/ptep/ptac150},
    url = {https://doi.org/10.1093/ptep/ptac150},
    eprint = {https://academic.oup.com/ptep/article-pdf/2023/4/042F01/50954267/ptac150.pdf},
}

@article{SimonsObservatory:2018koc,
    author = "Ade, Peter and others",
    collaboration = "Simons Observatory",
    title = "{The Simons Observatory: Science goals and forecasts}",
    eprint = "1808.07445",
    archivePrefix = "arXiv",
    primaryClass = "astro-ph.CO",
    doi = "10.1088/1475-7516/2019/02/056",
    journal = "JCAP",
    volume = "02",
    pages = "056",
    year = "2019"
}

@article{Ghigna2023,
  author  = {Ghigna, T. and Matsumura, T. and Sakurai, Y. and Takaku, R. and Komatsu, K. and Sugiyama, S. and Hoshino, Y. and Katayama, N.},
  title   = {Modelling TES non-linearity induced by a rotating HWP in a CMB polarimeter},
  journal = {Journal of Low Temperature Physics},
  year    = {2023},
  volume  = {211},
  number  = {5--6},
  pages   = {357--365},
  doi     = {10.1007/s10909-023-02939-5},
  eprint  = {2208.02952},
  archivePrefix = {arXiv},
  primaryClass  = {astro-ph.IM}
}

@article{DAlessandro2019,
  author  = {D'Alessandro, G. and Mele, L. and Columbro, F. and Pagano, L. and Piacentini, F. and de Bernardis, P. and Masi, S.},
  title   = {Systematic effects induced by half-wave plate precession into measurements of the cosmic microwave background polarization},
  journal = {Astronomy \& Astrophysics},
  year    = {2019},
  volume  = {627},
  pages   = {A160},
  doi     = {10.1051/0004-6361/201834495},
  eprint  = {1906.07010},
  archivePrefix = {arXiv},
  primaryClass  = {astro-ph.CO}
}

@article{Pancharatnam1955,
  author  = {Pancharatnam, S.},
  title   = {Achromatic combinations of birefringent plates},
  journal = {Proceedings of the Indian Academy of Sciences - Section A},
  year    = {1955},
  volume  = {41},
  pages   = {137--144},
  doi     = {10.1007/BF03047098}
}

@article{Hanany2005,
  author  = {Hanany, S. and Hubmayr, J. and Johnson, B. R. and Matsumura, T. and Oxley, P. and Thibodeau, M.},
  title   = {Millimeter-wave achromatic half-wave plate},
  journal = {Applied Optics},
  year    = {2005},
  volume  = {44},
  number  = {22},
  pages   = {4666--4670},
  doi     = {10.1364/AO.44.004666},
  eprint  = {physics/0503122},
  archivePrefix = {arXiv}
}

@article{Pisano2006,
  author  = {Pisano, Giampaolo and Savini, Giorgio and Ade, Peter A. R. and Haynes, Vic},
  title   = {Achromatic half-wave plate for submillimeter instruments in cosmic microwave background astronomy: experimental characterization},
  journal = {Applied Optics},
  year    = {2006},
  volume  = {45},
  number  = {27},
  pages   = {6982--6989},
  doi     = {10.1364/AO.45.006982}
}

@article{Matsumura2009,
  author  = {Matsumura, Tomotake and Hanany, Shaul and Ade, Peter and Johnson, Bradley R. and Jones, Terry J. and Jonnalagadda, Prashanth and Savini, Giorgio},
  title   = {Performance of three- and five-stack achromatic half-wave plates at millimeter wavelengths},
  journal = {Applied Optics},
  year    = {2009},
  volume  = {48},
  number  = {19},
  pages   = {3614--3625},
  doi     = {10.1364/AO.48.003614},
  eprint  = {0806.1518},
  archivePrefix = {arXiv},
  primaryClass = {astro-ph}
}

@article{Takaku:21,
author = {Ryota Takaku and Qi Wen and Scott Cray and Mark Devlin and Simon Dicker and Shaul Hanany and Takashi Hasebe and Teruhito Iida and Nobuhiko Katayama and Kuniaki Konishi and Makoto Kuwata-Gonokami and Tomotake Matsumura and Norikatsu Mio and Haruyuki Sakurai and Yuki Sakurai and Ryohei Yamada and Junji Yumoto},
journal = {Opt. Express},
number = {25},
pages = {41745--41765},
publisher = {Optica Publishing Group},
title = {Large diameter millimeter-wave low-pass filter made of alumina with laser ablated anti-reflection coating},
volume = {29},
month = {Dec},
year = {2021},
url = {https://opg.optica.org/oe/abstract.cfm?URI=oe-29-25-41745},
doi = {10.1364/OE.444848},
}

@article{Kusaka2018ABS,
    author  = {Kusaka, Akito and Appel, John and Essinger-Hileman, Thomas and others},
    title   = {Results from the Atacama B-mode Search (ABS) Experiment},
    journal = {Journal of Cosmology and Astroparticle Physics},
    volume  = {2018},
    number  = {09},
    pages   = {005},
    year    = {2018},
    doi     = {10.1088/1475-7516/2018/09/005}
}

@article{10.1093/mnras/stab317,
    author = {Duivenvoorden, Adriaan J and Adler, Alexandre E and Billi, Matteo and Dachlythra, Nadia and Gudmundsson, Jon E},
    title = {Probing frequency-dependent half-wave plate systematics for CMB experiments with full-sky beam convolution simulations},
    journal = {Monthly Notices of the Royal Astronomical Society},
    volume = {502},
    number = {3},
    pages = {4526-4539},
    year = {2021},
    month = {04},
    issn = {0035-8711},
    doi = {10.1093/mnras/stab317},
    url = {https://doi.org/10.1093/mnras/stab317},
    eprint = {https://academic.oup.com/mnras/article-pdf/502/3/4526/36386195/stab317.pdf},
}

\section{Details of the matrix formalism}
\label{appendixA}

For the $j$-th mode in the $i$-th layer, the electric-flux vector is constructed as
\begin{align}
    \bm{D}_{i2} &= \hat{\bm{k}}_{i2}\times\hat{\bm{\xi}},&
    \bm{D}_{i4} &= \hat{\bm{k}}_{i4}\times\hat{\bm{\xi}},\\
    \bm{D}_{i1} &= \hat{\bm{k}}_{i1}\times\bm{D}_{i2},&
    \bm{D}_{i3} &= \hat{\bm{k}}_{i3}\times\bm{D}_{i4}.
\end{align}
The corresponding electric-field vector and normalized magnetic-field vector are
\begin{align}
    \bm{\gamma}_{ij}
    &=
    \bm{\varepsilon}_{i}^{-1}\bm{D}_{ij},\\
    \bm{\zeta}_{ij}
    &=
    \frac{1}{n_{ij}}
    \hat{\bm{k}}_{ij}\times\bm{D}_{ij}.
\end{align}
The layer matrix is then given by the tangential components as
\begin{align}
    \bm{A}_i
    =
    \begin{pmatrix}
        \gamma_{i1,x} & \gamma_{i2,x} & \gamma_{i3,x} & \gamma_{i4,x}\\
        \gamma_{i1,y} & \gamma_{i2,y} & \gamma_{i3,y} & \gamma_{i4,y}\\
        \zeta_{i1,x} & \zeta_{i2,x} & \zeta_{i3,x} & \zeta_{i4,x}\\
        \zeta_{i1,y} & \zeta_{i2,y} & \zeta_{i3,y} & \zeta_{i4,y}
    \end{pmatrix}.
\end{align}

To extract the Jones transmission coefficients, we express the total transfer
matrix $\bm{M}$ defined in Eq.~\ref{eq:Mtot_sandbox} in the boundary-amplitude
basis $(E_t^p,E_r^p,E_t^s,E_r^s)$.



\end{document}